\documentclass[journal=aamick,manuscript=article,layout=traditional]{achemso}
\setkeys{acs}{etalmode=truncate,maxauthors=100}

\usepackage{chemformula} 
\usepackage[separate-uncertainty=true,separate-uncertainty,multi-part-units=single]{siunitx} 
\usepackage{booktabs} 
\usepackage{multirow} 
\usepackage{textgreek} 

\DeclareSIUnit\angstrom{\text {Å}} 

\newcommand{\vacCs}{$\mathrm{V_{\ch{Cs}}}$}
\newcommand{\vacI}{$\mathrm{V_{\ch{I}}}$}

\newcommand{\GBfirst}{$\mathrm{3 \Sigma (112) (0.4, 0)}$}
\newcommand{\GBsecond}{$\mathrm{5 \Sigma (210) (0.4, 0)}$}
\newcommand{\GBthird}{$\mathrm{3 \Sigma (111) (0, 0)}$}

\author{Mike Pols}
    \affiliation{Materials Simulation \& Modelling, Department of Applied Physics, Eindhoven University of Technology, 5600 MB, Eindhoven, The Netherlands}
    \alsoaffiliation{Center for Computational Energy Research, Department of Applied Physics, Eindhoven University of Technology, 5600 MB, Eindhoven, The Netherlands}
    \altaffiliation{Contributed equally to this work}
\author{Tobias Hilpert}
    \affiliation{Materials Simulation \& Modelling, Department of Applied Physics, Eindhoven University of Technology, 5600 MB, Eindhoven, The Netherlands}
    \alsoaffiliation{Laboratory of Inorganic Materials Chemistry, Schuit Institute of Catalysis, Department of Chemical Engineering and Chemistry, Eindhoven University of Technology, 5600 MB, Eindhoven, The Netherlands}
    \altaffiliation{Contributed equally to this work}
\author{Ivo A.W. Filot}
    \affiliation{Laboratory of Inorganic Materials Chemistry, Schuit Institute of Catalysis, Department of Chemical Engineering and Chemistry, Eindhoven University of Technology, 5600 MB, Eindhoven, The Netherlands}
    \alsoaffiliation{Center for Computational Energy Research, Department of Applied Physics, Eindhoven University of Technology, 5600 MB, Eindhoven, The Netherlands}
\author{Adri C.T. van Duin}
    \affiliation{Department of Mechanical Engineering, Pennsylvania State University, University Park, PA 16802, United States}
\author{Sof\'{i}a Calero}
    \affiliation{Materials Simulation \& Modelling, Department of Applied Physics, Eindhoven University of Technology, 5600 MB, Eindhoven, The Netherlands}
    \email{s.calero@tue.nl}
\author{Shuxia Tao}
    \affiliation{Materials Simulation \& Modelling, Department of Applied Physics, Eindhoven University of Technology, 5600 MB, Eindhoven, The Netherlands}
    \alsoaffiliation{Center for Computational Energy Research, Department of Applied Physics, Eindhoven University of Technology, 5600 MB, Eindhoven, The Netherlands}
    \email{s.x.tao@tue.nl}

\title{What Happens at Surfaces and Grain Boundaries of Halide Perovskites: Insights from Reactive Molecular Dynamics Simulations of \ch{CsPbI3}}


\begin{document}


\begin{abstract}

The commercialization of perovskite solar cells is hindered by the poor long-term stability of the metal halide perovskite (MHP) light absorbing layer. Solution processing, the common fabrication method for MHPs, produces polycrystalline films with a wide variety of defects, such as point defects, surfaces, and grain boundaries. Although the optoelectronic effects of such defects have been widely studied, the evaluation of their impact on the long-term stability remains challenging. In particular, an understanding of the dynamics of degradation reactions at the atomistic scale is lacking. In this work, using reactive force field (ReaxFF) molecular dynamics simulations, we investigate the effects of defects, in the forms of surfaces, surface defects and grain boundaries, on the stability of the inorganic halide perovskite \ch{CsPbI3}. Our simulations establish a stability trend for a variety of surfaces, which correlates well with the occurrence of these surfaces in experiments. We find that a perovskite surface degrades by progressively changing the local geometry of \ch{PbI_{x}} octahedra from corner- to edge- to face-sharing. Importantly, we find that \ch{Pb} dangling bonds and the lack of steric hindrance of \ch{I} species are two crucial factors that induce degradation reactions. Finally, we show that the stability of these surfaces can be modulated by adjusting their atomistic details, either by creating additional point defects or merging them to form grain boundaries. While in general additional defects, particularly when clustered, have a negative impact on the material stability, some grain boundaries have a stabilizing effect, primarily because of the additional steric hindrance.

\end{abstract}


\section{Introduction}

Metal halide perovskites (MHPs) with the \ch{ABX3} formula (\ch{A} = organic or inorganic cation; \ch{B} = metal cation; \ch{X} = halide anion) have attracted a great deal of attention as low-cost and high-performance semiconductors for applications in photovoltaics~\cite{kePerovskiteSolarCell2014, kimHighEfficiencyPerovskiteSolar2020} and light-emitting diodes~\cite{vanleRecentAdvancesHighEfficiency2018, shenEfficientPureBlue2022}. MHPs are typically synthesized using facile solution processing deposition techniques. Halide salts are mixed in solution with metal halide precursors, where solvent evaporation causes the formation of colloidal particles in solution, with further evaporation resulting in the growth of these particles in a polycrystalline perovskite film~\cite{yanHybridHalidePerovskite2015, nayakMechanismRapidGrowth2016, mcmeekinCrystallizationKineticsMorphology2017, flatkenSmallangleScatteringReveal2021}. This simple fabrication procedure offers a wide tunability in compositions and dimensions. However, it generally introduces a wide variety of defects in the material, ranging from point defects to crystal surfaces and grain boundaries~\cite{ballDefectsPerovskitehalidesTheir2016, onoReducingDetrimentalDefects2020}. For the majority of these defects it is found that they have limited electronic effect on the MHP, with point defects~\cite{kangHighDefectTolerance2017, chuSoftLatticeDefect2020, xueIntrinsicDefectsPrimary2022}, surfaces~\cite{haruyamaTerminationDependenceTetragonal2014, haruyamaSurfacePropertiesCH3NH3PbI32016} and grain boundaries~\cite{guoStructuralStabilitiesElectronic2017, shanSegregationNativeDefects2017, mckennaElectronicProperties1112018, caiAtomicallyResolvedElectrically2022} mainly resulting in electronically benign defect levels. While it is common knowledge that defects induce instability problems in MHPs, for instance through a defect-driven accumulation of defects at grain boundaries~\cite{parkAccumulationDeepTraps2019, phungRoleGrainBoundaries2020, digirolamoIonMigrationInducedAmorphization2020} or material degradation at perovskite interfaces~\cite{guoDegradationMechanismsPerovskite2021}, the understanding of the dynamics of the degradation reactions at an atomistic scale and their impact on the long-term stability of the materials and devices is limited.

Until now, the common understanding in the literature has been that during the degradation of MHPs the material disintegrates into its precursors. Specifically, various experiments have demonstrated the formation of \ch{PbX2} and amorphous \ch{PbX_{2-x}} (with \ch{X} = \ch{I} or \ch{Br}) using X-ray diffraction (XRD) measurements in the degradation of MHPs~\cite{coningsIntrinsicThermalInstability2015, fanLayerbyLayerDegradationMethylammonium2017, yangComprehensiveUnderstandingHeatinduced2018, albertiPbClusteringPbI22019, luDecisiveInfluenceAmorphous2021}. Moreover, metallic lead has also been found as a degradation product in MHPs~\cite{albertiPbClusteringPbI22019, luDecisiveInfluenceAmorphous2021}. Using electron microscopy it has been established that the degradation of halide perovskites tends to occur at surfaces and grain boundaries~\cite{albertiPbClusteringPbI22019, manekkathodiObservationStructuralPhase2020, luCompositionalInterfaceEngineering2020}. For that reason, a variety of efforts have been devoted to minimizing the occurrence of such grain boundaries in perovskite films to enhance the stability of the materials and devices~\cite{yenLargeGrainedPerovskite2016, seoIonicLiquidControl2016}. However, despite this knowledge on the perovskite films and the decomposition products, a detailed understanding of the degradation pathways of halide perovskites is still lacking.

In this work, our objective is to provide insights into the role of surfaces and grain boundaries in the long-term stability of MHPs. To do so, we perform reactive molecular dynamics simulations, using a reactive force field (ReaxFF) we developed for \ch{CsPbI3}~\cite{polsAtomisticInsightsDegradation2021}. This ReaxFF force field makes use of a dynamical bond order to describe the breaking and creation of bonds~\cite{vanduinReaxFFReactiveForce2001, chenowethDevelopmentApplicationReaxFF2008, senftleReaxFFReactiveForcefield2016}. The simulations allow us to characterize the evolution of the atomic species under thermal stress and to establish a stability trend of a variety of surfaces. Furthermore, based on the simulation trajectories, we establish what structural features make a surface stable or unstable, and in case of decomposition, through which atomistic processes it proceeds. Finally, we find that additional point defects are generally detrimental to perovskite stability, whereas grain boundaries can have either positive or negative effects on the stability of the perovskite lattice.

\section{Results and Discussion}

In the following sections, we present our findings on the stability of the surfaces of inorganic \ch{CsPbI3}. In section~\ref{sec:structural_models} we show the structural models of the surfaces and grain boundaries that we investigate in this work. We specifically highlight the equivalence between orthorhombic and cubic surface models. In section~\ref{sec:phase_stability_surfaces}, before we show the defect-induced chemical instability, we first illustrate the effects of surfaces on the phase stability of \ch{CsPbI3} by comparing the structural details of the surfaces with those in the bulk. Section~\ref{sec:surface_stability} evaluates the stability of different perovskite surfaces under thermal stress and elaborates on the mechanism of defect-induced degradation. We then explore the effects of additional defects located at surfaces, by analyzing the stability of the surfaces of perovskites with defects in section~\ref{sec:surface_defect_effects}. Finally, in section~\ref{sec:grain_boundaries}, we investigate the effects of grain boundaries on the stability of the perovskite lattice.

\subsection{Structural models} \label{sec:structural_models}

\begin{figure*}[htbp!]
    \includegraphics{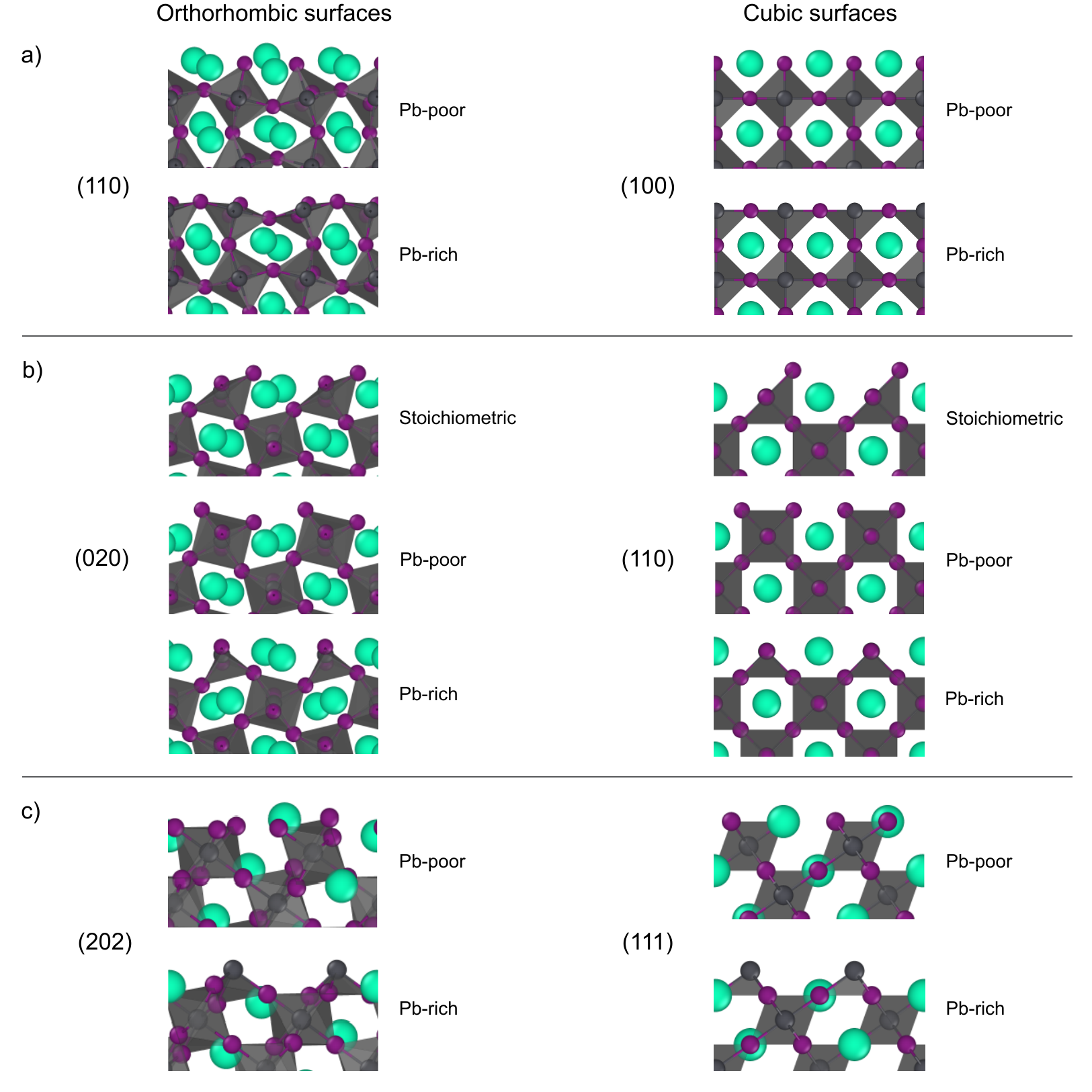}
    \caption{\label{surface_terminations} Structural models of the different surface orientations for orthorhombic \ch{CsPbI3} perovskite slabs. (a) (110) / (100), (b) (020) / (110) and (c) (202) / (111) surfaces are shown for the orthorhombic (left) / cubic (right) phases. For each of the slabs the exposed surface is shown at the top.}
\end{figure*}

To model the surfaces, we use slab models, shown in Figure~\ref{surface_terminations}a-c. These include \ch{CsPbI3} slabs with the (110), (020) and (202) planes of orthorhombic \ch{CsPbI3} exposed. Of these surfaces, the (110) and (020) planes of orthorhombic \ch{CsPbI3} are those most prevalent in XRD experiments~\cite{suttonCubicOrthorhombicRevealing2018}. Despite the limited occurrence of the orthorhombic (202) plane in experiments, we include a slab with this orientation for completeness. This allows for the investigation of slabs with corners, edges, and faces of the \ch{PbI_{x}} octahedra exposed, which is the case for the (110), (020) and (202) orthorhombic surfaces, respectively. Additionally, each surface is created with varying terminations. We differentiate between the following slab terminations: stoichiometric, \ch{Pb}-poor, \ch{Pb}-rich, an explanation for these names can be found in Supporting Note 1. We emphasize that although we here refer to the slabs and their exposed planes in the orthorhombic form, all slabs attain time-averaged cubic structures at elevated temperatures due to thermal fluctuations. For comparison, the equivalent time-averaged cubic structure found when heating each orthorhombic surface is also shown in Figure~\ref{surface_terminations}. Additional details of the surface models can be found in Supporting Note 1.

A variety of \ch{CsPbI3} grain boundaries is studied to find their effects on the stability of halide perovskites. We focus on the \GBfirst{}, \GBsecond{} and \GBthird{} grain boundaries, which are created from the cubic phase of \ch{CsPbI3}~\cite{guoStructuralStabilitiesElectronic2017}. We specifically investigate these grain boundary models because they do not contain unsaturated atoms with dangling bonds, the effects of which are probed with the above surface models. Additional details on the creation and naming of the grain boundaries can be found in Supporting Note 2.

\subsection{Phase stability near surface} \label{sec:phase_stability_surfaces}

\begin{figure*}[htbp!]
    \includegraphics{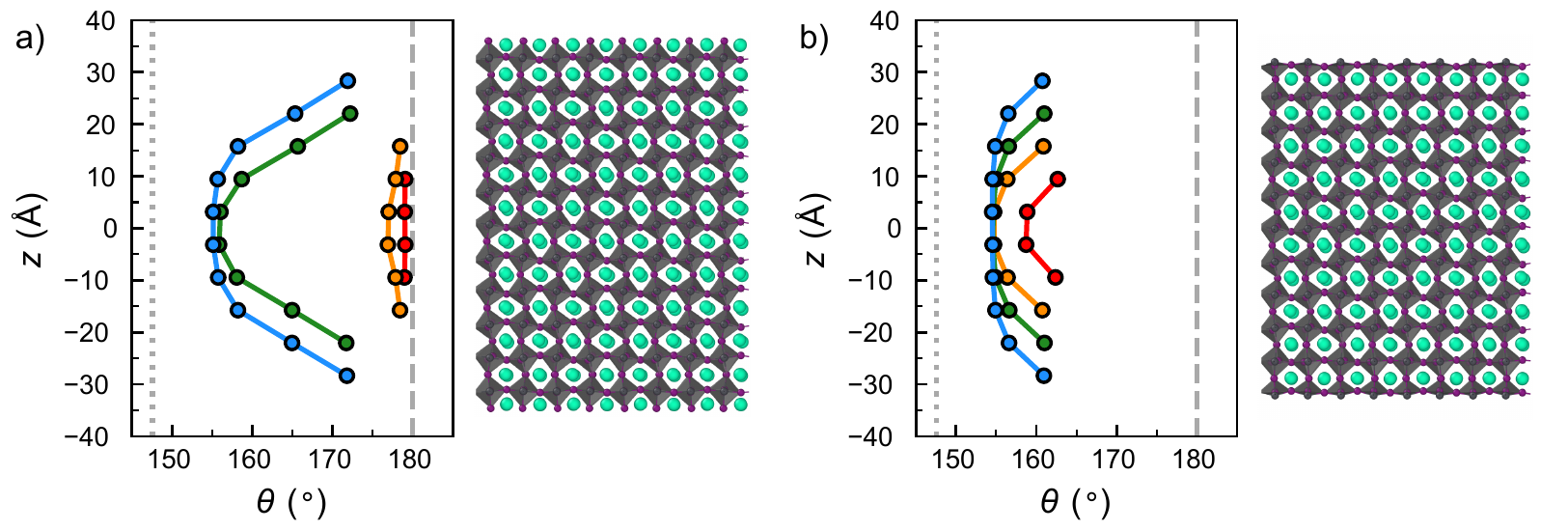}
    \caption{\label{surface_effects} Time-averaged values of the \ch{Pb}-\ch{I}-\ch{Pb} valence angle ($\theta$) perpendicular to the surfaces grouped together along the depth of the slabs, with $z = 0$ indicating the center of the slab for the (a) \ch{Pb}-poor and (b) \ch{Pb}-rich terminations. The red, orange, green and blue lines indicate the slabs of 4, 6, 8 and 10 octahedral cages thick, respectively. The time-averaged structure of the 10 layer slabs of \ch{CsPbI3} are shown next to the graphs. The values of the valence angle as obtained from density functional theory (DFT) calculations for the orthorhombic (\SI{148}{\degree}) and cubic (\SI{180}{\degree}) phases are indicated with the dotted and dashed lines, respectively.}
\end{figure*}

To assess the effects of surfaces on the bulk structure, we simulate \ch{CsPbI3} slabs of the most commonly encountered (110) orthorhombic surface~\cite{suttonCubicOrthorhombicRevealing2018}, which is equivalent to the (100) cubic surface~\cite{eperonInorganicCaesiumLead2015, haqueEffectsHydroiodicAcid2018, liuModifyingSurfaceTermination2021}. Slabs with a thickness of 4, 6, 8 or 10 octahedral cages and a \ch{Pb}-poor or \ch{Pb}-rich termination are simulated at a constant temperature of \SI{300}{\K}. We use the \ch{Pb}-\ch{I}-\ch{Pb} valence angle ($\theta$) to probe the effects of the surfaces on the bulk structure of the perovskite during the simulation. The time-averaged values of the \ch{Pb}-\ch{I}-\ch{Pb} angles oriented in the direction perpendicular to the surfaces are shown in Figure~\ref{surface_effects}a (\ch{Pb}-poor) and Figure~\ref{surface_effects}b (\ch{Pb}-rich), together with the structural model of the slab with 10 layers. The structural models of slabs with a smaller thickness can be found in Supporting Note 3.

We find that regardless of the termination, the \ch{Pb}-\ch{I}-\ch{Pb} valence angle near the center of the perovskite slab has an on average lower value ($\theta_{\mathrm{MD}} = \SI{155}{\degree}$) than close to the surface ($\theta_{\mathrm{MD}} = 160-175\SI{}{\degree}$). A comparison of these valence angles to those found for bulk structures using density functional theory (DFT) shows us the center of the slab is orthorhombic ($\theta_{\mathrm{DFT}} = \SI{148}{\degree}$), while the regions near the surface tend to exhibit a more cubic structure ($\theta_{\mathrm{DFT}} = \SI{180}{\degree}$). We hypothesize that small perovskite domains attain a core-shell structure, in which the core has an orthorhombic structure and the surface-layer is cubic-like, which we highlight could impact the optoelectronic properties of such domains~\cite{marronnierAnharmonicityDisorderBlack2018}. Notably, this agrees with experimental findings for \ch{CsPbI3}, where progressively larger nanocrystals undergo a phase transition from a cubic-like phase to an orthorhombic phase~\cite{zhaoSizeDependentLatticeStructure2020}. We find this cubic-like surface layer is finite, approximately \SI{2}{\nm} thick, which might be a slight overestimation that stems from a small underestimation of the phase transition temperatures in ReaxFF~\cite{polsAtomisticInsightsDegradation2021}. It has been shown that the large octahedral distortion in the orthorhombic phase of \ch{CsPbI3} results in poor \ch{Cs}-\ch{I} contacts in the material, which destabilizes the lattice, whereas the longer \ch{Cs}-\ch{I} contacts in cubic \ch{CsPbI3} result in a stabilization of the perovskite~\cite{strausUnderstandingInstabilityHalide2020}. Therefore, we propose that this core-shell structure could explain the enhanced stability of nanostructured \ch{CsPbI3} in experiments~\cite{shpatzdayanEnhancingStabilityPhotostability2018, zhaoSizeDependentLatticeStructure2020} against the conversion into the nonperovskite yellow phase~\cite{stoumposSemiconductingTinLead2013, marronnierAnharmonicityDisorderBlack2018}.

\subsection{Surface stability} \label{sec:surface_stability}

To investigate the thermal stability of perovskite surfaces, we subject a collection of perovskite slabs (Figure~\ref{surface_terminations}) to different temperatures ranging from \SI{300}{\K} to \SI{700}{\K} in steps of \SI{50}{\K}. From these simulations we determine the onset temperature of material degradation, which we use as an indication for the stability of the perovskite surfaces. In the definition of material degradation we include all processes that result in a crystal lattice that deviates from the pristine form, which include but are not limited to the formation of defect complexes, the clustering of atomic species and the breakaway of atoms from the surface. The onset temperatures for the degradation of all slabs can be found in Table~\ref{onset_temperatures}. 

\begin{table}[htbp!]
    \caption{\label{onset_temperatures} Onset temperatures for lattice degradation of perovskite slabs with a different orientations and terminations. For each surface the octahedral feature that protrudes the surfaces is indicated. The (*) denotes a stable surface up to and including \SI{700}{\K} and (-) the absence of such a surface in the investigations.}
    \begin{tabular}{ccccc}
        \toprule
        Orientations        & Surface feature   & Stoichiometric    & \ch{Pb}-poor      & \ch{Pb}-rich      \\ \midrule
        (110)               & Corner            & -                 & *                 & \SI{550}{\K}      \\
        (020)               & Edge              & \SI{550}{\K}      & \SI{300}{\K}      & \SI{300}{\K}      \\
        (202)               & Face              & -                 & \SI{400}{\K}      & \SI{300}{\K}      \\ \bottomrule
    \end{tabular}
\end{table}

We find that (110) orthorhombic slabs possess the highest resistance to thermal stress. The \ch{Pb}-poor surface is stable for temperatures up to \SI{700}{\K} for simulations up to \SI{5}{\ns}, with the \ch{Pb}-rich surface showing decomposition of the lattice from \SI{550}{\K} and higher. This stability matches the trend in surface formation energies observed for the equivalent cubic surfaces in the ReaxFF force field validation (see Supporting Note 4). A lower thermal stability is observed for the (020) orthorhombic surface. The stoichiometric (020) surface of \ch{CsPbI3} is stable up to \SI{450}{\K} with the rearrangement of surface iodine atoms occurring at \SI{500}{\K} (details in Supporting Note 5) and degradation of the surface occurring at \SI{550}{\K} and higher. Both the \ch{Pb}-rich and \ch{Pb}-poor termination of the (020) orthorhombic surface are unstable, exhibiting the clustering of atomic species near the surface at \SI{300}{\K}. Finally, the (202) orthorhombic surface is the least stable, with the \ch{Pb}-poor termination degrading from \SI{400}{\K} and up and the \ch{Pb}-rich termination already decomposing at \SI{300}{\K}. On the basis of these results, we rank the surface orientations from most stable to least stable as: (110) $>$ (020) $>$ (202). We emphasize that this observed stability trend correlates well with the occurrence of these surfaces in XRD experiments, in which the most stable surfaces appear predominantly~\cite{suttonCubicOrthorhombicRevealing2018, eperonInorganicCaesiumLead2015, haqueEffectsHydroiodicAcid2018, liuModifyingSurfaceTermination2021}.

\begin{figure*}[htbp!]
    \includegraphics{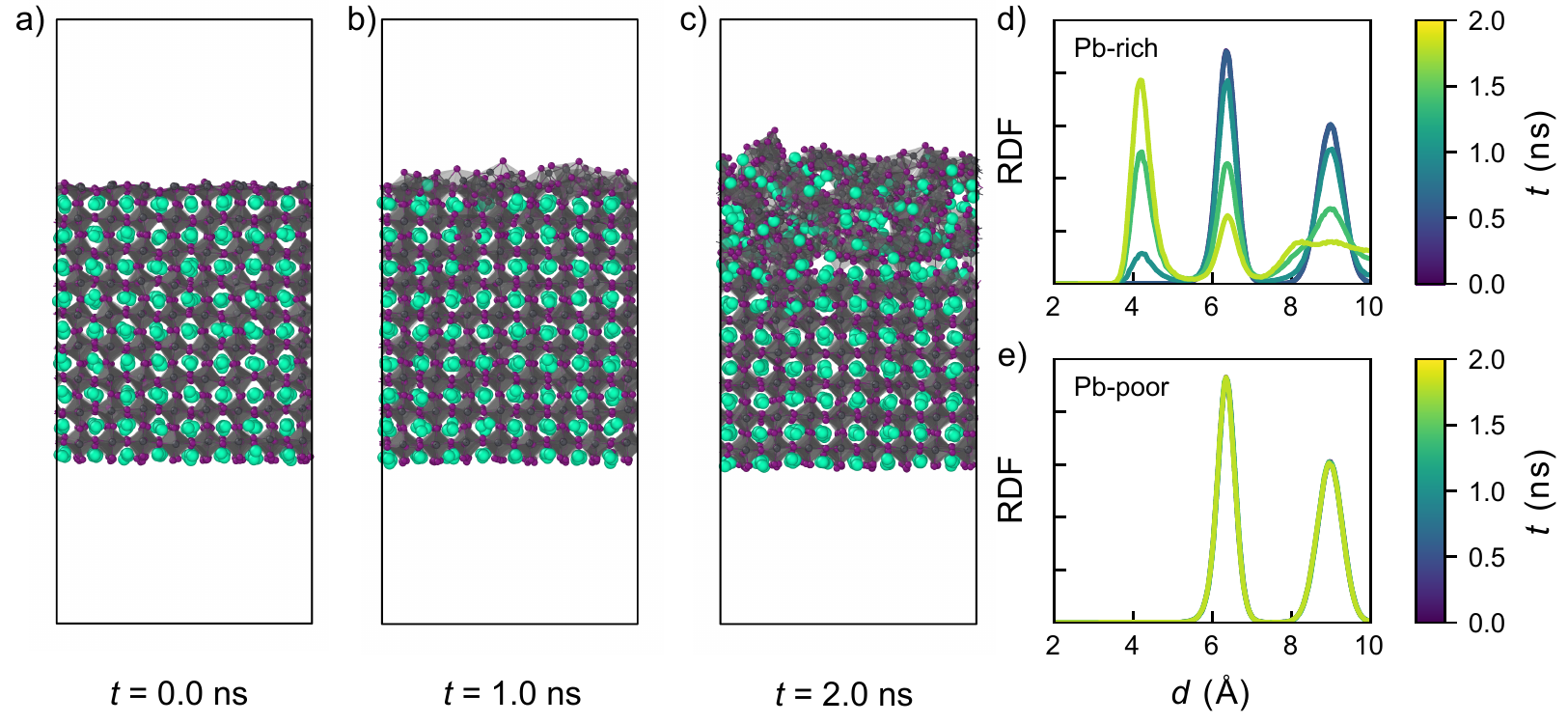}
    \caption{\label{degradation_dynamics} Degradation of a (110) orthorhombic perovskite surface at \SI{600}{\K}, with the \ch{Pb}-rich surface on the top and the \ch{Pb}-poor surface at the bottom of the slab. (a-c) Structural snapshots of the degrading perovskite slab. (d-e) Time evolution of the \ch{Pb}-\ch{Pb} radial distribution function (RDF) for the \ch{Pb}-rich and \ch{Pb}-poor surfaces of the perovskite slab, demonstrating that the degradation of the \ch{Pb}-rich surface starts from a clustering of \ch{Pb} species.}
\end{figure*}

For more insights into the degradation dynamics, we take a closer look at the degradation of a (110) orthorhombic slab simulated at \SI{600}{\K}. In Figure~\ref{degradation_dynamics}a-c we show snapshots of the lattice degradation during the simulation. Consistent with our earlier observations, the snapshots show that under these conditions the (110) \ch{Pb}-poor surface remains intact, while the \ch{Pb}-rich surface shows decomposition, leading to the formation of a \ch{Pb_{x}I_{y}} complex. To quantify the degradation processes, we analyze the radial distribution functions (RDFs) between the \ch{Pb}-\ch{Pb} atom pairs in the simulated system in Figure~\ref{degradation_dynamics}d-e and for the remaining atom pairs in Supporting Note 6. On the \ch{Pb}-rich surface, the \ch{Pb}-\ch{Pb} RDF shows a simultaneous increase in the peak at \SI{4.2} {\angstrom} and a decrease of the peak at \SI{6.3} {\angstrom}, the timing of which coincides with the onset of degradation seen in Figure~\ref{degradation_dynamics}a-c. On the contrary, the \ch{Pb}-poor surface does not show any change in peak intensity and therefore shows no signs of decomposition. Based on these observations, we conclude that the degradation of the perovskite lattice can be characterized by the clustering of \ch{Pb} atoms, which has been observed in experiments at perovskite grain boundaries before~\cite{albertiPbClusteringPbI22019}. By comparing the position of the emerging \ch{Pb}-\ch{Pb} peak (\SI{4.2}{\angstrom}) to the interatomic distances of \ch{Pb} species from DFT in layered \ch{PbI2} (\SI{4.55}{\angstrom}) or yellow phase \ch{CsPbI3} (\SI{4.65}{\angstrom}), we can conclude that the decomposition product is not a pure form of either of these two material phases, but a more amorphous \ch{Pb_{x}I_{y}} domain. The formation of these domains matches observations from several experimental reports in which the formation of \ch{PbI2} domains is observed in electron microscopy experiments at grain boundaries~\cite{albertiPbClusteringPbI22019, manekkathodiObservationStructuralPhase2020, luDecisiveInfluenceAmorphous2021}.

\begin{figure*}[htbp!]
    \includegraphics{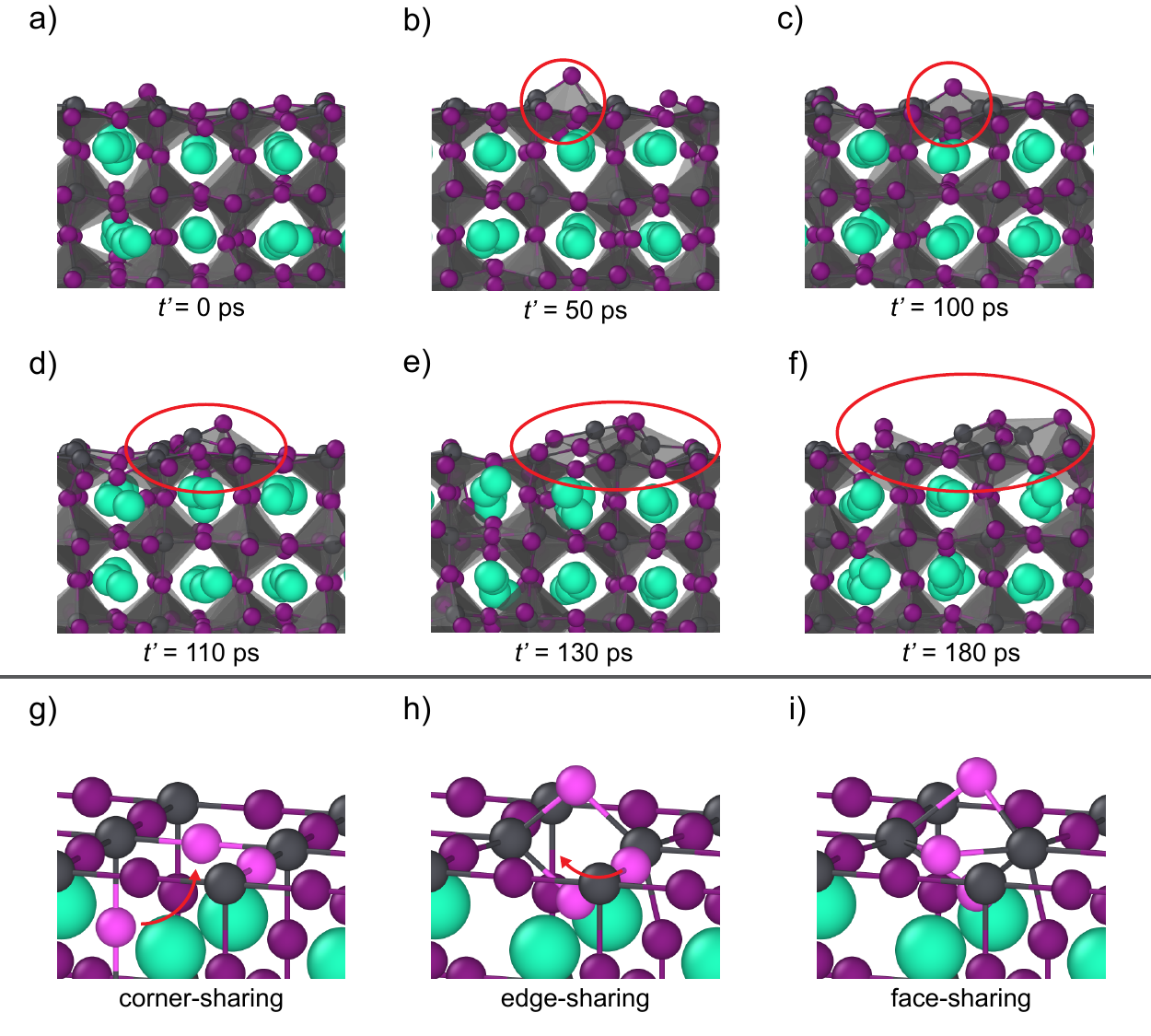}
    \caption{\label{degradation_mechanism} Snapshots of the degradation of a \ch{Pb}-rich (110) orthorhombic perovskite surface. (a) Non-degraded perovskite surface. (b-c) A relatively long-lived iodine Frenkel defect at the perovskite surface, resulting in an edge-sharing complex. (d-e) A face-sharing complex at the surface that starts to break away from the surface. (f) Decomposed perovskite lattice at the surface. (g-i) Schematic representation of the critical steps in the degradation mechanism where neighboring \ch{PbI_{x}} octahedra change from corner- to edge- to face-sharing. The species highlighted in pink are the \ch{I} atoms involved in the formation of the surface defect. The figure makes use of a shifted time axis, with $t' = \SI{0}{\ps}$ corresponding to $t = \SI{0.75}{\ns}$ in simulation time.}
\end{figure*}

From all the decomposing \ch{CsPbI3} surfaces reported above, we observe that all surfaces degrade through a similar mechanism. To illustrate the key steps of this mechanism, we show the degradation of the \ch{Pb}-rich orthorhombic (110) surface at \SI{600} {\K} as an example in Figure~\ref{degradation_mechanism}. In the first step of the degradation process, an iodine Frenkel defect is formed in the perovskite lattice (Figure~\ref{degradation_mechanism}a-c). As a result of this, two \ch{PbI_{x}} octahedra form an edge-sharing complex, as opposed to the corner-sharing geometry in a regular perovskite lattice. Here, two \ch{Pb} atoms are bound together by two \ch{I} atoms. This edge-sharing complex acts as a metastable state in our simulations, exhibiting lifetimes of up to \SI{50}{\ps}. In the next step, an additional \ch{I} atom is added to the edge-sharing complex, transforming it into a face-sharing complex, in which the interatomic distance of the \ch{Pb} atoms involved in the complex significantly decreases (Figure~\ref{degradation_mechanism}d). We regard the formation of this face-sharing complex as the starting point of the degradation of the surface. Shortly after its formation, this face-sharing complex breaks away from the surface, initiating the decomposition of the perovskite lattice near the surface (Figure~\ref{degradation_mechanism}e-f). Altogether we find that temperature affects the rate at which the degradation reaction proceeds, with more significant thermal fluctuations at elevated temperatures resulting in a faster degradation of the perovskite lattice.

We highlight that the critical steps of the degradation mechanism of perovskite surfaces (Figure~\ref{degradation_mechanism}g-i) resemble the degradation near iodine vacancies in the bulk of \ch{CsPbI3}~\cite{polsAtomisticInsightsDegradation2021}. In this previous study we also found that iodine interstitials do not initiate the degradation of perovskites. Thus, we establish that Frenkel defects, which have previously mainly been connected to ion migration~\cite{phungRoleGrainBoundaries2020, meggiolaroFormationSurfaceDefects2019}, are a main cause of lattice instabilities that originate from the vacancy part of the interstitial-vacancy defect pair. Based on the above we note that two distinct features make perovskite surfaces more prone to degradation: 1) an abundance of dangling bonds and 2) a lack of steric hindrance. Using DFT calculations it has been shown that the density and type of dangling bonds affects the formation energies of perovskite surfaces~\cite{wangDensityFunctionalStudies2015}. Here, we posit that such dangling bonds impact not only the thermodynamic stability of these surfaces but also the dynamical stability. In particular, we find that undercoordinated \ch{Pb} species readily form new bonds that then progressively degrade the halide perovskite. Additionally, the stability trend of the perovskite surfaces found earlier can be classified according to the protruding surface features as: corner $>$ edge $>$ face. We stress that in this ranking the \ch{PbI_{x}} octahedra, and specifically the \ch{I} species, experience an increasingly smaller steric hindrance from the \ch{Cs} species at the surface, allowing for the movement of these octahedra and thus a larger tendency for perovskite degradation. Finally, these two factors also explain the high thermal stability of the \ch{Pb}-poor (110) orthorhombic surface. In addition to the absence of \ch{Pb} dangling bonds, the \ch{Cs} atoms also sterically hinder the movement of the undercoordinated \ch{I} species at the surface. Together, this inhibits the decomposition, making the surface very stable.

\subsection{Effect of additional point defects on surfaces} \label{sec:surface_defect_effects}

To assess the effects of additional point defects on the stability of \ch{CsPbI3} perovskite surfaces, we use the \ch{Pb}-poor (110) orthorhombic surface as a model system because this perovskite surface has shown high thermal stability (Section~\ref{sec:surface_stability}). Since the surface is dominated by \ch{Cs} and \ch{I} species, with theoretical support for the occurrence of vacancies of both species~\cite{longEffectSurfaceIntrinsic2019}, we look at the effect of such vacancies at \SI{600}{\K}, individually and when they are clustered. In the remainder of this work we refer to the \ch{Cs} and \ch{I} vacancies as \vacCs{} and \vacI{}. The structural models used for the investigation of these defects can be found in Supporting Note 7.

\begin{figure*}[htbp!]
    \includegraphics{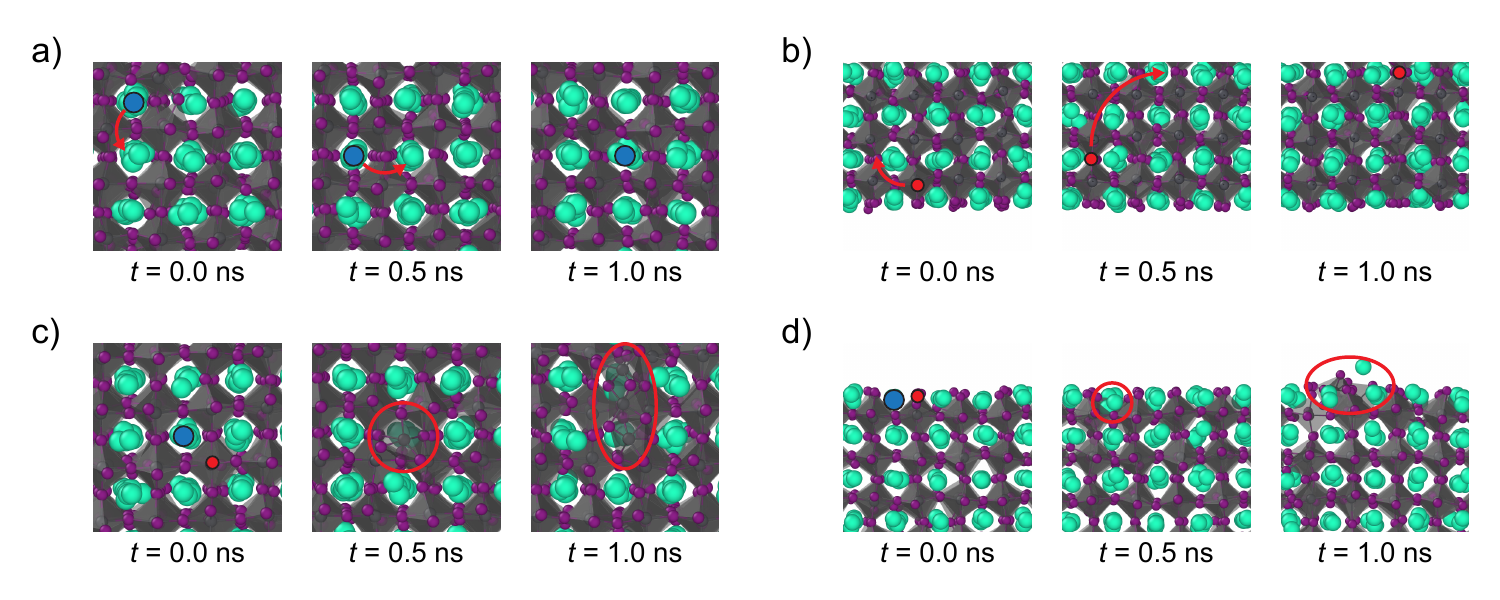}
    \caption{\label{defect_effects} Snapshots of different point defects in \ch{CsPbI3} slabs. (a) Top view of a surface with a \vacCs{} defect, showing the motion of the defect across the surface. (b) Side view of a surface with a \vacI{} defect, showing the migration of the defect away from the surface into the bulk. (c-d) A top and side view of a perovskite surface with a \vacCs{} and \vacI{} defect pair, showing the degradation of the perovskite surface caused by the presence of this defect at the surface. The \vacCs{} and \vacI{} defects are indicated with a blue and red sphere, respectively.}
\end{figure*}

In their isolated form, the defects have relatively benign effects on the lattice stability, which appears very similar to that observed for bulk \ch{CsPbI3}~\cite{polsAtomisticInsightsDegradation2021}, but when the defects form pairs, it is found that they do significantly impact the stability of the perovskite lattice. The \vacCs{} defect remains on the surface of the slab, where it migrates across the surface as shown in Figure~\ref{defect_effects}a, without resulting in the degradation of the lattice. In contrast, the motion of \vacI{} is not bound to the surface of the perovskite. As shown in Figure~\ref{defect_effects}b, a \vacI{} defect can migrate into and out of the bulk of the perovskite. Moreover, in some cases the \vacI{} defect causes the perovskite surface to degrade; however, this process is not restricted to perovskite surfaces and also occurs in the bulk of inorganic perovskites~\cite{polsAtomisticInsightsDegradation2021}. In Figure~\ref{defect_effects}c-d the effects of a closely spaced defect pair of \vacI{} and \vacCs{} are shown. During simulation, we observe that this defect cluster tends to stay together at a fixed position on the surface, only occasionally splitting into isolated \vacCs{} and \vacI{} defects, which indicates an enhancement of the defect trapping ability of surfaces for halide defects~\cite{caddeoDominantRoleSurfaces2020}, specifically when paired with cation vacancies. When the defect pair remains clustered, the pair is particularly detrimental for the stability of the perovskite lattice. Specifically, we observe \ch{Pb} species readily moving away from their original position the lattice in close proximity to the defect pair ($t = \SI{0.5}{\ns}$ in Figure~\ref{defect_effects}), initiating the degradation of the lattice by allowing \ch{Pb} and \ch{I} species to cluster and form an amorphous \ch{Pb_{x}I_{y}} domain. We connect the tendency for a defect pair to act as a degradation center to the earlier-established factors for perovskite degradation: the presence of \ch{Pb} dangling bonds and limited steric hindrance for \ch{I} species near the defect pair.

\subsection{Grain boundaries} \label{sec:grain_boundaries}

\begin{figure*}[htbp!]
    \includegraphics{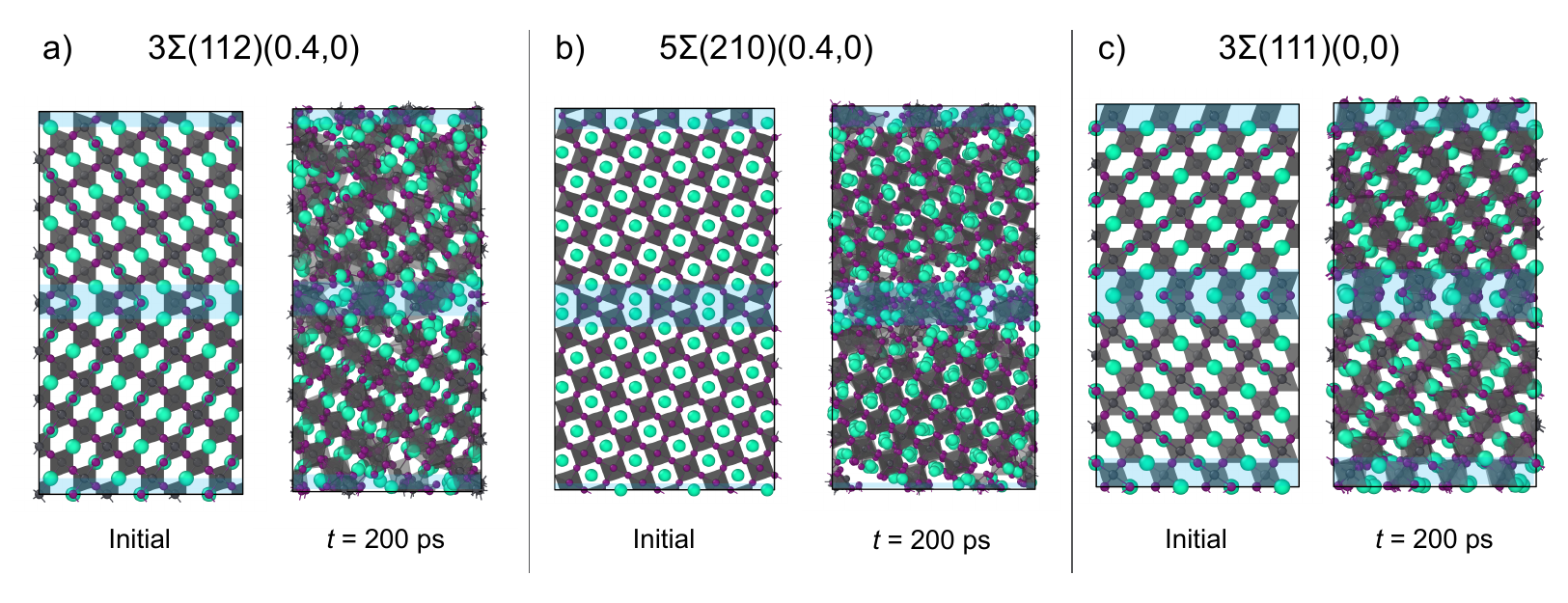}
    \caption{\label{gb_evolution} Dynamical evolution of the \ch{CsPbI3} grain boundaries at \SI{600}{\K} after \SI{200}{\ps} from their initial structure. (a) \GBfirst{} grain boundary. (b) \GBsecond{} grain boundary. (c) \GBthird{} grain boundary. The blue areas highlight the grain boundaries in the structures.}
\end{figure*}

We assess the stability of grain boundaries from simulations at \SI{600}{\K}. The structure of the model systems after \SI{200}{\ps} of simulation is shown in Figure~\ref{gb_evolution}. The \GBfirst{} (Figure~\ref{gb_evolution}a) and \GBsecond{} (Figure~\ref{gb_evolution}b) grain boundaries exhibit clustering of atomic species in the grain boundary region. This clustering results in the formation of amorphous \ch{Pb_{x}I_{y}} domains, an observation that is consistent with the surfaces presented above (Section~\ref{sec:surface_stability}) and the experimentally observed degradation of perovskites at grain boundaries~\cite{manekkathodiObservationStructuralPhase2020, luDecisiveInfluenceAmorphous2021, albertiPbClusteringPbI22019}.  Contrary to the other two grain boundaries, the \GBthird{} grain boundary (Figure~\ref{gb_evolution}c), also known as a twinning plane~\cite{mckennaElectronicProperties1112018}, does not show any degradation throughout the duration of the simulation (\SI{2}{\ns}).

\begin{figure*}[htbp!]
    \includegraphics{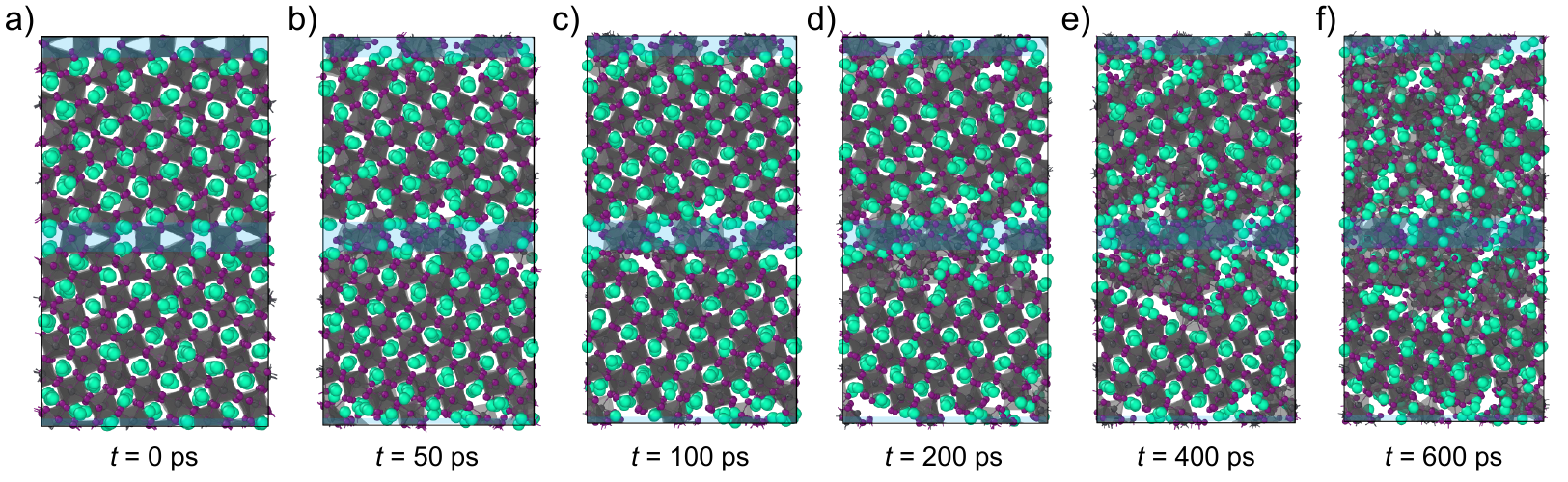}
    \caption{\label{gb_mechanism} Degradation of a \GBsecond{} grain boundary. (a-c) The degradation initiates at the grain boundary and (d-f) proceeds into the bulk of the perovskite. The blue areas in the figures highlight the grain boundary.}
\end{figure*}

To investigate the degradation mechanism of these grain boundaries in more detail, we look into the time evolution of the degradation \GBsecond{} grain boundary (Figure~\ref{gb_mechanism}), which for the \GBfirst{} and \GBthird{} grain boundaries is shown in Supporting Note 8. Careful inspections point to a general mechanism in which the degradation is initiated by the movement of iodine atoms near the grain boundary, resulting in the formation of small \ch{Pb_{x}I_{y}} domains at the grain boundary (Figure~\ref{gb_mechanism}a-b). Similar to perovskite surfaces, we find that these small \ch{Pb_{x}I_{y}} domains grow progressively larger (Figure~\ref{gb_mechanism}c), resulting in the degradation progressing into the perovskite bulk (Figure~\ref{gb_mechanism}d-f). Due to an absence of any dangling bonds in the grain boundary models, we can connect this observed material instability to the lack of steric hindrance at grain boundaries. In particular, both the \GBfirst{} and \GBsecond{} grain boundaries lack \ch{Cs} species that can block the clustering of closely spaced \ch{Pb} and \ch{I} species, making them unstable. In contrast, the grain boundary \GBthird{} has intact face-sharing \ch{PbI_{x}} octahedra at the grain boundary with cavity filling \ch{Cs} species that sterically hinder the movement of these octahedra. Altogether this stabilizes the \GBthird{} grain boundary, making this grain boundary more stable than its (111) cubic surface analog, which, based on its equivalent (202) orthorhombic surface, is unstable from \SI{400}{\K} onward.

\section{Conclusion}

In summary, using a ReaxFF force field, we study structural and thermal stability effects of surfaces and grain boundaries in the inorganic halide perovskite \ch{CsPbI3}. We show that surfaces affect the crystal phase close to the surface, which attains a cubic-like structure that is approximately \SI{2}{\nm} in size. We believe that this surface region is responsible for the enhanced structural stability of nanostructured \ch{CsPbI3}. Under a thermal stress that ranges from \SI{300}{\K} to \SI{700}{\K}, we find a stability trend for orthorhombic \ch{CsPbI3} of: (110) $>$ (020) $>$ (202). This trend matches the occurrence of these surfaces in experiments, in which the most stable surfaces are most predominantly observed. Comparing all investigated structures, we propose two important factors that are responsible for the degradation of perovskites: 1) the presence of dangling bonds, particularly for \ch{Pb} species and 2) a lack of steric hindrance, especially for \ch{I} species. These two factors explain the high thermal stability observed for the \ch{Pb}-poor (110) orthorhombic surface. The surface only has \ch{I} dangling bonds, no \ch{Pb} dangling bonds, and the motion of the \ch{I} species is blocked by the \ch{Cs} species on the surface, resulting in a stable surface. All other surfaces, which do not satisfy these two conditions, eventually decompose through a common mechanism. In this mechanism first an iodine Frenkel defect is formed in close proximity to the surface; after some time this defect grows to form a complex of two face-sharing \ch{PbI_{x}} octahedra; finally the complex breaks away from the surface, growing larger by accumulating more \ch{Pb} and \ch{I} species, leading to the formation of a large amorphous \ch{Pb_{x}I_{y}} domain on the surface.

The stability of aforementioned surfaces deteriorates when additional point defects (\vacI{} and \vacCs{}) are created on the surfaces, which results in the formation of \ch{Pb} dangling bonds and a local decrease of steric hindrance. The latter factor also explains the stability of the investigated grain boundaries. For example, although the \GBfirst{} and \GBsecond{} grain boundaries do not contain any dangling bonds, the lack of steric hindrance for \ch{Pb} and \ch{I} species facilitates the clustering of these species, leading to the degradation of the material. On the contrary, the \GBthird{} grain boundary contains cavity filling \ch{Cs} species that sterically hinder the typical degradation-inducing movement of the face-sharing octahedra, making it stable up to \SI{600}{\K}.

Based on the above, we propose that strategies to stabilize halide perovskites can include the following aspects: i) passivating defects, primarily those leading to unsaturated bonds, i.e. halide vacancies, through passivating agents such as halogens like \ch{F}~\cite{liCationAnionImmobilization2019} and \ch{Cl}~\cite{gaoHalideExchangePassivation2022} or carbonyl~\cite{xieDecouplingEffectsDefects2021} and azo~\cite{maStablePerovskiteSolar2021} containing ligands; ii) the grafting of surfaces with sterically hindering groups, such as phenylalkylammonium~\cite{guoPhenylalkylammoniumPassivationEnables2021} and even bulkier organic groups~\cite{wuEfficientStableCsPbI32019}; (iii) optimizing the synthesis conditions, through the type of precursors, solvents, additives etc., to stimulate the growth of non-detrimental surface orientations~\cite{foleyControllingNucleationGrowth2016}, reduce the formation of grain boundaries~\cite{peteramalathasControlledGrowthLarge2020} and suppress the formation of defects altogether~\cite{yangDefectSuppressionOriented2021}.

\section{Computational details}

\subsection{Molecular dynamics}

ReaxFF molecular dynamics simulations were performed in AMS2021~\cite{AMS2021}. All ReaxFF simulations were done using the earlier developed \ch{CsPbI3} ReaxFF force field~\cite{polsAtomisticInsightsDegradation2021}, a validation of this force field for \ch{CsPbI3} surfaces can be found in Supporting Note 4. Before the molecular dynamics runs, all structural models were optimized with the ReaxFF force field. The dynamical simulations used a simulation timestep of \SI{0.25}{\fs}. The thermostat and barostat use a damping constant of $\tau_{T} = \SI{100}{\fs}$ and $\tau_{p} = \SI{2500}{\fs}$, respectively. Whenever slabs were simulated, the vacuum layer used was at least \SI{50}{\angstrom} in the $z$-direction. In the case of slabs, the barostat was only allowed to scale the non-vacuum directions of the model system ($x$- and $y$-directions). The initial velocities of the particles were assigned according to a Maxwell-Boltzmann distribution of the initial temperature. Simulation snapshots and structural models were all visualized using OVITO~\cite{stukowskiVisualizationAnalysisAtomistic2009}.

In the simulations investigating the structural effects of surfaces on the bulk structure of perovskites, we equilibrated the systems to the target temperature of \SI{300}{\K} in an NPT-ensemble for \SI{200}{\ps}. During the equilibration stage we employed a Berendsen thermostat and Berendsen barostat~\cite{berendsenMolecularDynamicsCoupling1984} to control the temperature and pressure, respectively. Production runs were started from the final frame of the equilibration run and took \SI{200}{\ps}. In the production runs, the temperature and pressure were controlled with a NHC-thermostat~\cite{martynaNoseHooverChains1992} with chain length of 10 and an MTK barostat~\cite{martynaConstantPressureMolecular1994}. The time-averaged values of the \ch{Pb}-\ch{I}-\ch{Pb} valence angles were extracted from the time-averaged structure we obtained by averaging the atomic positions over the full duration of the \SI{200}{\ps} production simulations.

The stability of the perovskite surfaces and grain boundaries was investigated using a three-step approach. In the first two steps, the equilibration stage, we used a Berendsen thermostat and Berendsen barostat to control the temperature and pressure. The first step was used to slowly heat the system from \SI{300}{\K} to the desired target temperature during \SI{100}{\ps}, with the second step maintaining the system at its constant target temperature for \SI{100}{\ps}. The full equilibration of the system was run in the non-reactive mode of ReaxFF in AMS2021, in which the bonds can only be updated but not newly formed, to prevent unwanted reactions during the equilibration of the system. The final stage, the production simulation, was run in an NPT-ensemble for which the starting point was the final frame of the equilibration. The temperature and pressure were controlled, respectively, with an NHC-thermostat with a chain length of 10 and an MTK-barostat. Each of the production runs was \SI{2}{\ns} long, except in the assessment of the stability of the \ch{Pb}-poor (110) orthorhombic surface for which we used \SI{5}{\ns} long simulations.

\subsection{Density Functional Theory}

Density functional theory (DFT) calculations were performed with the projector augmented wave (PAW) method as implemented in the Vienna Ab-Initio Simulation Package (VASP)~\cite{kresseInitioMoleculardynamicsSimulation1994, kresseEfficiencyAbinitioTotal1996, kresseEfficientIterativeSchemes1996, kresseUltrasoftPseudopotentialsProjector1999}.  The electron exchange-correlation interaction was described using the Perdew, Burke and Ernzerhof (PBE) functional~\cite{perdewGeneralizedGradientApproximation1996} with long-range dispersive interactions accounted for by the DFT-D3(BJ) dispersion correction~\cite{grimmeEffectDampingFunction2011}. We treated the outermost electrons of \ch{Cs} (5s\textsuperscript{2}5p\textsuperscript{6}6s\textsuperscript{1}); \ch{Pb} (5d\textsuperscript{10}6s\textsuperscript{2}6p\textsuperscript{2}) and \ch{I} (5s\textsuperscript{2}5p\textsuperscript{5}) as valence electrons with the plane wave basis set expanded to an energy cutoff of \SI{500}{\eV}.

The geometries were optimized by allowing all: the ionic positions, cell shape and cell volume to change until convergence of \SI{1E-3}{\meV} and \SI{10}{\meV\per\angstrom} was reached in energy and forces, respectively. The Brillouin zones were sampled using a Monkhorst-Pack mesh~\cite{monkhorstSpecialPointsBrillouinzone1976}, with the following $k$-space grids resulting in energy convergence to within \SI{1}{\meV}/atom: \ch{CsI}: $12 \times 12 \times 12$; \ch{PbI2}: $11 \times 11 \times 7$; cubic \ch{CsPbI3}: $10 \times 10 \times 10$; orthorhombic \ch{CsPbI3}: $7 \times 7 \times 5$; yellow phase \ch{CsPbI3}: $13 \times 6 \times 4$.


\begin{acknowledgement}

S.T. acknowledges funding by the Computational Sciences for Energy Research (CSER) tenure track program of Shell and NWO (Project No. 15CST04-2) and NWO START-UP support from the Netherlands.

\end{acknowledgement}


\begin{suppinfo}

Structural models of the surfaces; structural models of grain boundaries; time-averaged structures of 4, 6 and 8 octahedral layer thick perovskite slabs; ReaxFF force field validation using surface energies; illustration of the rearrangement of surface iodine for the stoichiometric (020) orthorhombic surface; explanation of RDFs and overview of the RDFs for remaining atom pairs; structural models of the defective surfaces; snapshots of the evolution of the \GBfirst{} and \GBthird{} grain boundaries

Structure files of the bulk perovskite and precursor structures, the pristine and defective perovskite slabs, and the perovskite grain boundaries.

Simulation trajectories at 600 K of the (110) orthorhombic perovskite slab during degradation, (110) orthorhombic Pb-poor perovskite slabs with isolated and paired \vacCs{} and \vacI{} defects, and the time evolution of the \GBfirst{}, \GBsecond{} and \GBthird{} perovskite grain boundaries.

\end{suppinfo}


\bibliography{ms}

\end{document}


\clearpage

\tableofcontents

\clearpage


\section*{Supporting Notes}

\subsection{Structural models for surfaces}

\subsubsection{Surface termination names}

The different surface terminations are named according to the relative concentration of species at these surfaces. To determine the surface concentration of the species, we create perovskite slabs with the same termination at both ends and consequently calculate the ratios of the atoms that make up these slabs. An overview of the ratios found for the different slabs is shown in Table~\ref{surface_ratios}. The naming scheme is based on the relative ratio of the \ch{I} and \ch{Pb} species at each surface, resulting in the following names: stoichiometric (\ch{I}:\ch{Pb} $=$ 3), \ch{Pb}-poor (\ch{I}:\ch{Pb} $>$ 3) and \ch{Pb}-rich (\ch{I}:\ch{Pb} $<$ 3).

\begin{table}[htbp!]
    \caption{\label{surface_ratios} Relative ratio of the occurrence of atomic species as determined for perovskite slabs with the same termination at both sides of the perovskite slab.}
    \begin{tabular}{ccccc}
        \toprule
        Orientation             & Termination       & \ch{Pb}:\ch{Cs}   & \ch{I}:\ch{Cs}    & \ch{I}:\ch{Pb}    \\ \midrule
        \multirow{2}{*}{(110)}  & \ch{Pb}-poor      & 0.89              & 2.78              & \textbf{3.13}     \\
                                & \ch{Pb}-rich      & 1.13              & 3.25              & \textbf{2.89}     \\ \midrule
        \multirow{3}{*}{(020)}  & Stoichiometric    & 1.00              & 3.00              & \textbf{3.00}     \\
                                & \ch{Pb}-poor      & 1.00              & 3.15              & \textbf{3.15}     \\
                                & \ch{Pb}-rich      & 1.00              & 2.86              & \textbf{2.86}     \\ \midrule
        \multirow{2}{*}{(202)}  & \ch{Pb}-poor      & 0.94              & 3.00              & \textbf{3.20}     \\
                                & \ch{Pb}-rich      & 1.07              & 3.00              & \textbf{2.81}     \\ \bottomrule
    \end{tabular}
\end{table}

\clearpage

\subsubsection{Surface models}

\begin{figure*}[htbp!]
    \includegraphics{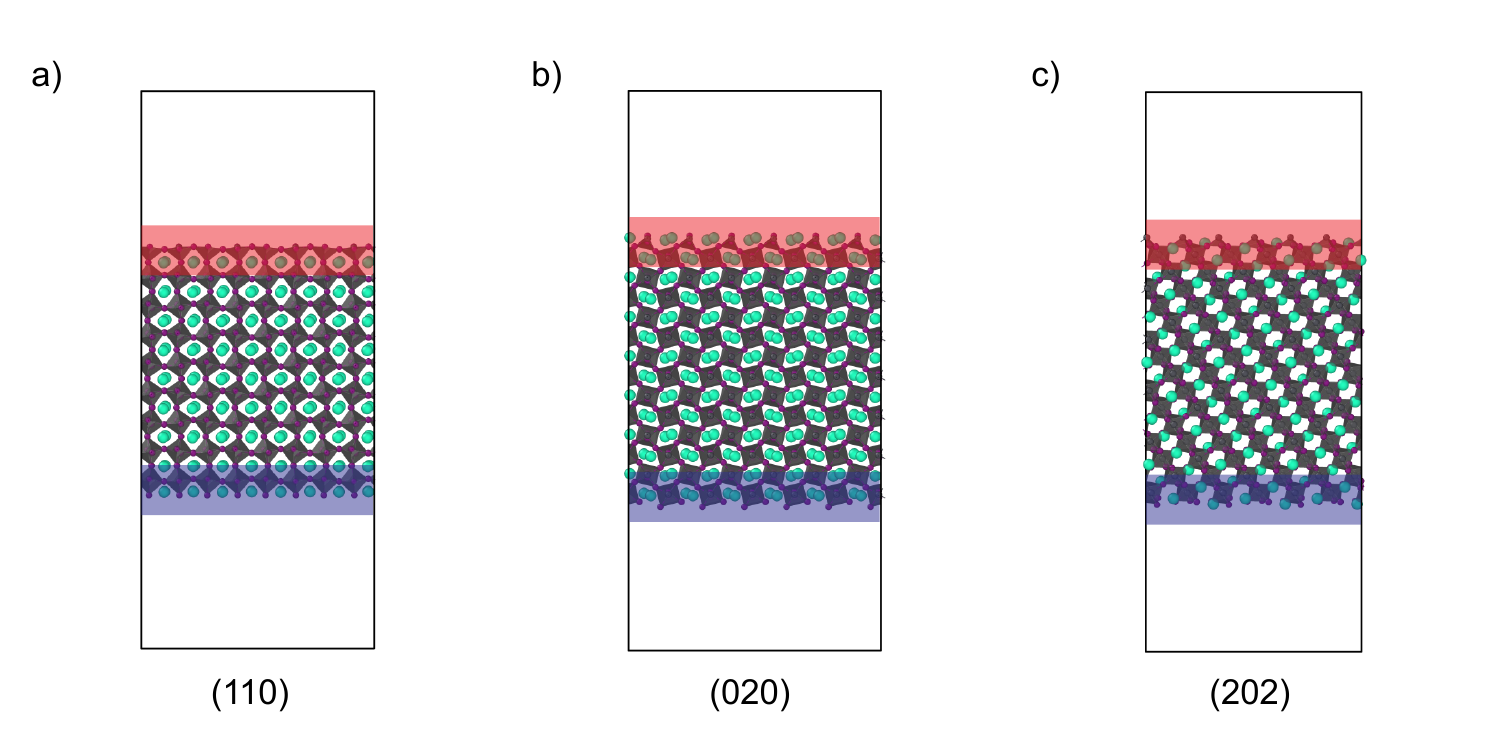}
    \caption{\label{surface_models} Structural models of the \ch{Pb}-poor surfaces as created from orthorhombic \ch{CsPbI3}. (a) (110) orthorhombic surface. (b) (020) orthorhombic surface. (c) (202) orthorhombic surface. The \ch{Pb}-rich and \ch{Pb}-poor terminated surfaces are indicated with the red and blue color, respectively.}
\end{figure*}

In this work the surface models are created in AMS2021~\cite{AMS2021}. An orthorhombic unit cell is used as the starting point for the surface models. Each model slab is at least \SI{50}{\angstrom} thick, with the lateral dimension spanning at least \SI{45}{\angstrom}. An overview of the full structural models of the doubly terminated (\ch{Pb}-poor and \ch{Pb}-rich) slabs is shown in Figure~\ref{surface_models}.

\clearpage

\subsection{Structural models for grain boundaries}

\subsubsection{Grain boundary names}

The names of the grain boundaries follow from the work of Guo et al.~\cite{guoStructuralStabilitiesElectronic2017}. The grain boundaries in this work are created with the coincidence site lattice (CSL) model. In this approach, two grains are tilted with a certain angle, until their surface planes coincide. An example of such a grain boundary structure is the \GBfirst{} grain boundary. In this naming scheme the $3$ in front of $\Sigma$ denotes that every third point on the surface planes coincides. The number that follows $\Sigma$, $(112)$, denotes the lattice plane along which the intersection occurs. The final two numbers indicate the overall fractional shift of the grains with respect to each other. For $(0.4,0)$ this means that one of the grains is shifted along the $[100]$ direction by $40\%$ of the length of the unit cell vector and without any shift along the $[010]$ direction. Note these shift directions hold because the grain boundaries are formed perpendicular to the $z$-axis.

\clearpage

\subsubsection{Grain boundary models}

\begin{figure*}[htbp!]
    \includegraphics{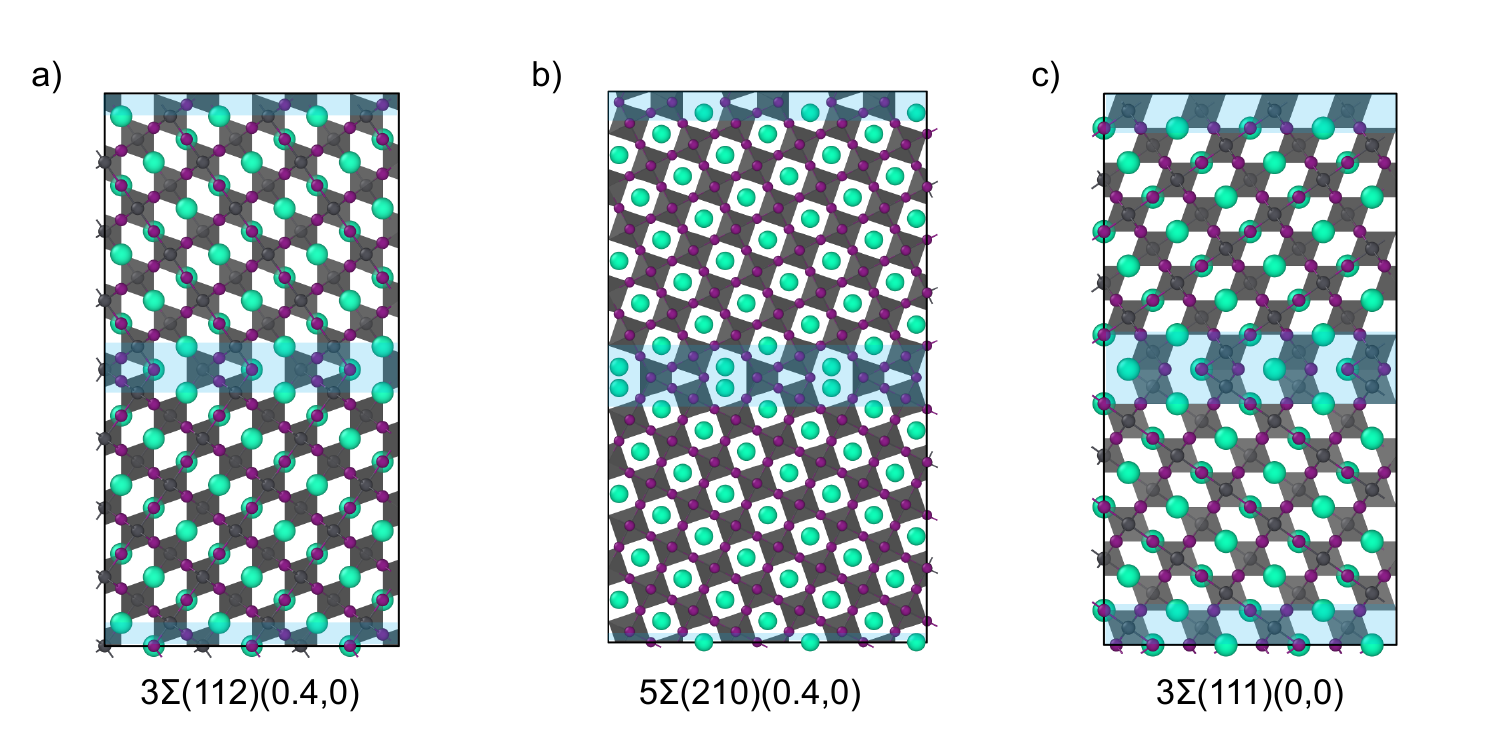}
    \caption{\label{gb_models} Structural models of the grain boundaries as created from cubic \ch{CsPbI3}. (a) \GBfirst{} grain boundary. (b) \GBsecond{} grain boundary. (c) \GBthird{} grain boundary. The blue areas highlight the grain boundaries in the structures.}
\end{figure*}

The grain boundary models in this work are created with the \texttt{aimsgb} package~\cite{chengAimsgbAlgorithmOpensource2018}. A cubic \ch{CsPbI3} unit cell is used as the starting point for the grain boundary models. An overview of the initial structures of the grain boundary models is shown in Figure~\ref{gb_models}. In each model the grain boundaries are separated by at least \SI{25}{\angstrom}. The lateral dimensions of the grain boundary models are at least \SI{30}{\angstrom} in size.

\clearpage

\subsection{Structural effects of surfaces}

\begin{figure*}[htbp!]
    \includegraphics{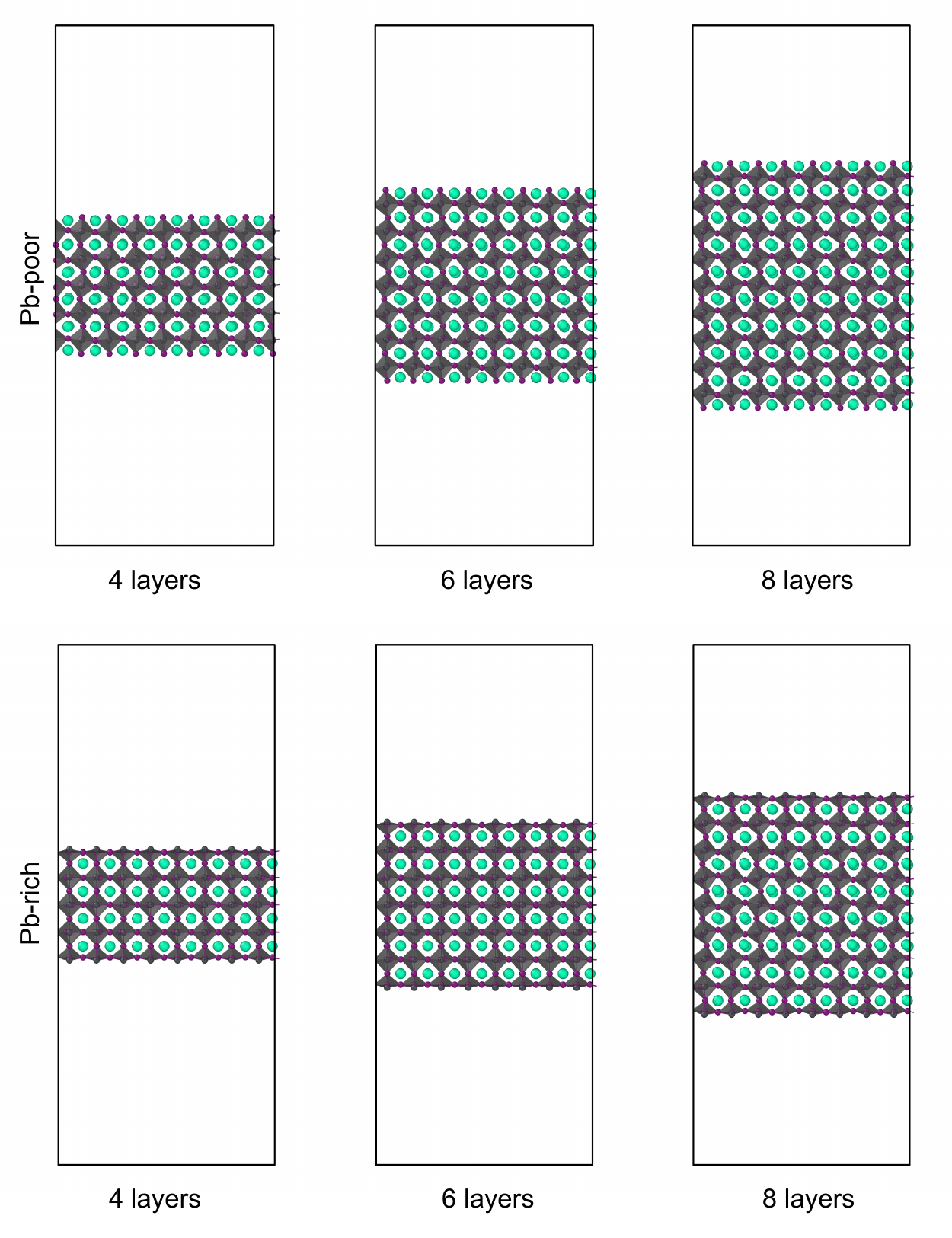}
    \caption{\label{averaged_slabs} Time-averaged perovskite slab structures for both \ch{Pb}-poor and \ch{Pb}-rich slabs of 4, 6 and 8 layers of octahedral cages.}
\end{figure*}

The time-averaged structures of the 10 layer thick slabs of \ch{CsPbI3} are shown in Figure 2 in the manuscript. For a complete overview of the slabs with varying thickness, the time-averaged structural models of the slabs of 4, 6 and 8 octahedral cages thick are shown in Figure~\ref{averaged_slabs}, which demonstrates that thinner slabs attain more cubic-like structures.

\clearpage

\subsection{Force field validation}

\begin{figure*}[htbp!]
    \includegraphics{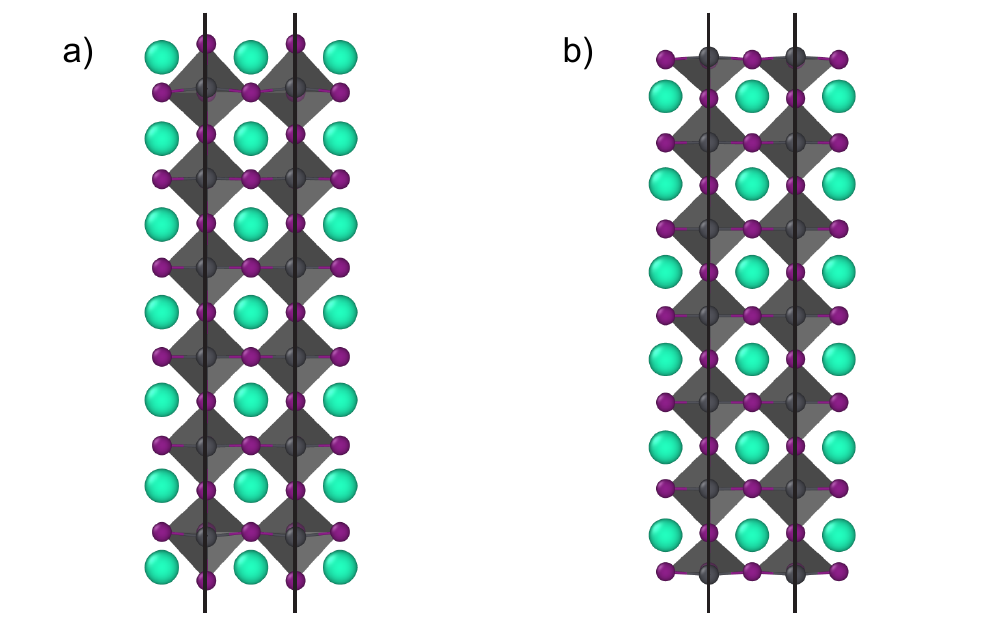}
    \caption{\label{validation_structures} Structural models of the slabs used in the validation of the force field. (a) \ch{Pb}-poor terminated slab. (b) \ch{Pb}-rich terminated slab. The simulation cell is indicated with the black lines.}
\end{figure*}

Slab geometries were created from optimized bulk structures of the cubic phase of \ch{CsPbI3} in both DFT and ReaxFF. During relaxation of the slabs, only the ionic positions were allowed to change until convergence was obtained in energy and forces. In the DFT calculations a $6 \times 6 \times 1$ $k$-points mesh was used during slab optimizations. At least \SI{15}{\angstrom} of vacuum was included to minimize the interaction between periodic images of the slab. The structural models of the perovskite slabs are shown in Figure~\ref{validation_structures}.

Surface energies ($\gamma$) were calculated as
\begin{equation}
    \gamma = \frac{E_{\mathrm{slab}} - \left( n \cdot E_{\ch{CsPbI3}} \pm E_{\mathrm{precursor}} \right) }{2 A}
\end{equation}
where $E_{\mathrm{slab}}$ is the energy of the perovskite slab, $n$ the number of formal units of \ch{CsPbI3}, $E_{\mathrm{\ch{CsPbI3}}}$ is the energy of cubic \ch{CsPbI3}, $E_{\ch{precursor}}$ is the energy of the perovskite precursor and $A$ the surface area of the perovskite slab. To make use of a common point of reference for the validation, we pick \ch{CsI} as the precursor, resulting in $n = 6$ and an addition of the precursor energy ($+ E_{\mathrm{precursor}}$) for the \ch{Pb}-poor terminated slab and $n = 7$ and a subtraction of the precursor energy ($- E_{\mathrm{precursor}}$) for the \ch{Pb}-rich terminated slab.

An overview of the surface areas and calculated surface energies of the perovskite slabs with the \ch{Pb}-rich and \ch{Pb}-poor termination is shown in Table~\ref{validation_values}. From this comparison we find that the used ReaxFF force field matches the stability trend of our DFT simulations, as well as those found in literature~\cite{seiduAtomicElectronicStructure2021, longEffectSurfaceIntrinsic2019}. In both computational methods the \ch{Pb}-poor surface termination is found to be significantly more stable than the \ch{Pb}-rich surface termination. Although the relative magnitude for DFT and ReaxFF does not exactly match, the comparison validates that the force field is transferable for the simulation of surfaces.

\begin{table}[htbp!]
    \caption{\label{validation_values} The areas ($A$) and surface energies ($\gamma$) for the different \ch{CsPbI3} slab terminations as calculated with DFT and ReaxFF.}
    \begin{tabular}{cccc}
        \toprule
        Surface termination     & $A$ ($ \SI{}{\angstrom\squared}$)  & $\gamma_{\mathrm{DFT}}$ ($\SI{}{\joule\per\meter\squared}$) & $\gamma_{\mathrm{ReaxFF}}$  ($\SI{}{\joule\per\meter\squared}$) \\ \midrule
        \ch{Pb}-poor            & 39.58                             & 0.095                                                     & -0.087                        \\
        \ch{Pb}-rich            & 39.58                             & 0.222                                                     & 0.144                         \\ \bottomrule
    \end{tabular}
\end{table}

\clearpage

\subsection{Surface rearrangements of iodine}

\begin{figure*}[htbp!]
    \includegraphics{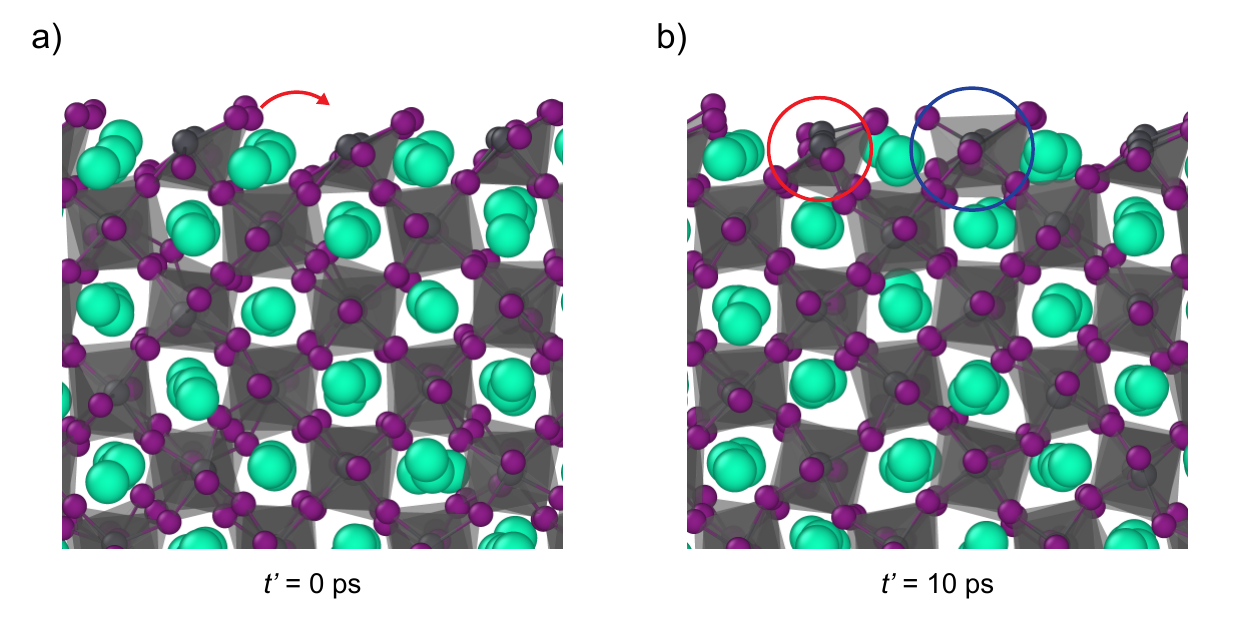}
    \caption{\label{surface_rearrangements} Snapshots of the rearrangement of surface iodine atoms for the orthorhombic (020) stoichiometric surface at \SI{500}{\K}. (a) Stuctural arrangement before the migration of surface iodine. (b) Surface geometry after surface iodine rearrangment, with the blue and red circle indicating a \ch{Pb}-poor and \ch{Pb}-rich octahedra, respectively. A shifted time axis $t'$ is used.}
\end{figure*}

For the stoichiometric orthorhombic (020) surface it was observed that iodine species would migrate across the surface from \SI{500}{\K} and up. To illustrate this behavior, some simulation snapshots from \SI{500}{\K} simulations are shown in Figure~\ref{surface_rearrangements}. We underline that the jump of an iodine atom to a neighboring \ch{Pb} atom (Figure~\ref{surface_rearrangements}a) forms local \ch{Pb}-rich and \ch{Pb}-poor octahedra, which become the degradation centers at elevated temperatures.

\clearpage

\subsection{Radial distribution functions}

The RDF between the atoms of the species $a$ and $b$ is calculated using
\begin{equation}
    g_{ab} \left( r \right) = \frac{1}{N_{a} N_{b}} \sum^{N_{a}}_{i = 1} \sum^{N_{b}}_{j = 1} \langle \delta \left( \lvert \vec{r}_{i} - \vec{r}_{j}  \rvert - r \right) \rangle,
\end{equation}
which sums over Dirac delta functions $\delta$ that are offset by the interatomic distance between the $i$th atom of species $a$ located at $\vec{r}_{i}$ and the $j$th atom of species $b$ located at $\vec{r}_{j}$. The quantity is averaged over the full length of the simulation indicated by the ensemble averaging ($\langle \cdots \rangle$).

\begin{figure*}[htbp!]
    \includegraphics{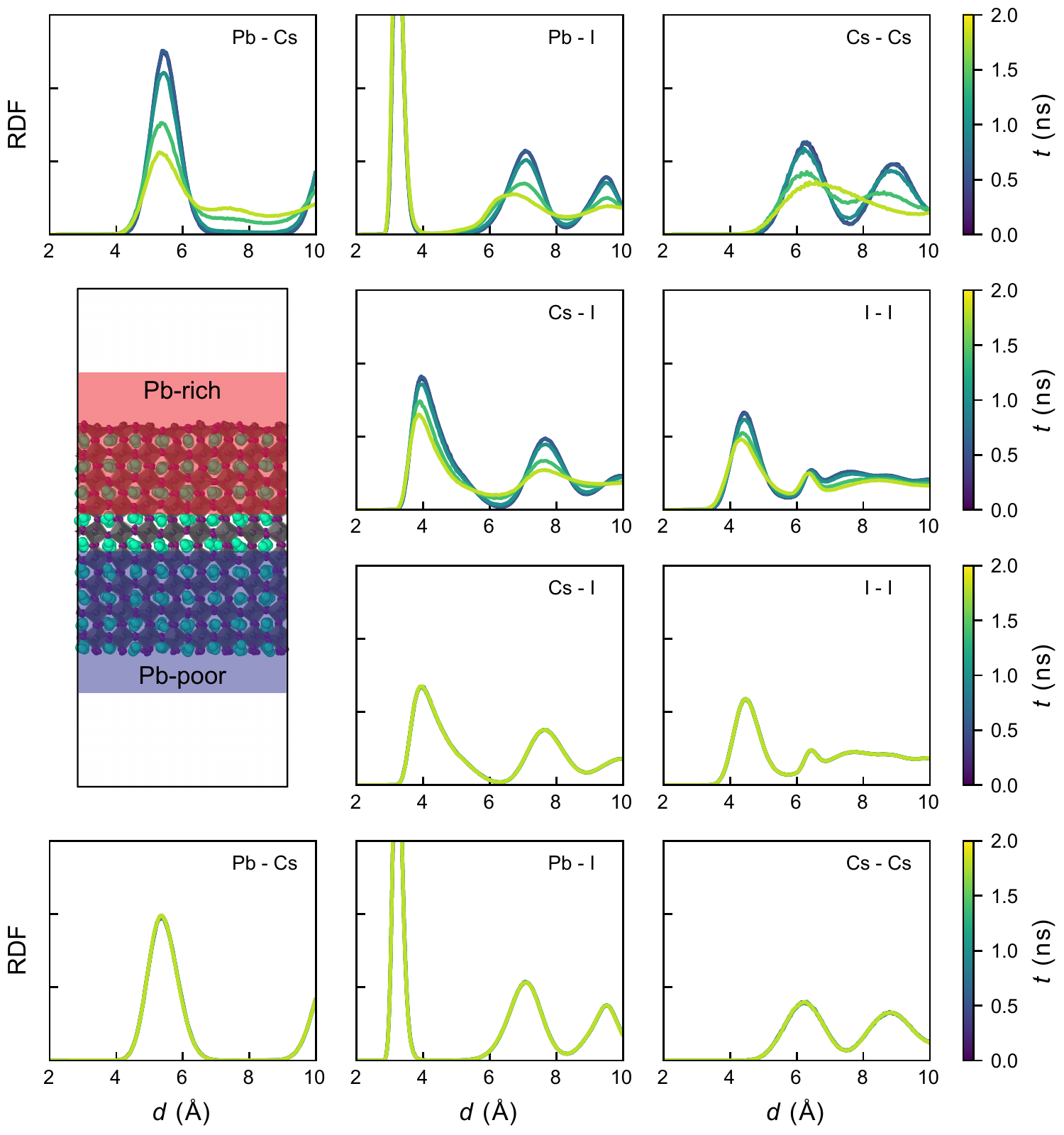}
    \caption{\label{rdfs_combined} Time-resolved radial distribution functions (RDFs) for the (110) surface of orthorhombic \ch{CsPbI3}. The figure shows the RDFs of the \ch{Pb}-\ch{Pb}, \ch{Pb}-\ch{Cs}, \ch{Pb}-\ch{I}, \ch{Cs}-\ch{Cs}, \ch{Cs}-\ch{I} and \ch{I}-\ch{I} atom pairs. The RDFs in the top half correspond to the \ch{Pb}-rich surface (red) and the figures at the bottom half to the \ch{Pb}-poor surface (blue).}
\end{figure*}

In this work, we employ the radial distribution function (RDF) as implemented in the \texttt{MDAnalysis} package~\cite{michaud-agrawalMDAnalysisToolkitAnalysis2011, gowersMDAnalysisPythonPackage2016}. Surface specific RDFs are obtained by only including those atoms that are within 4 layers of \ch{Pb} atoms from the surface. The time information is obtained by dividing the trajectory of the degradation simulation into five blocks of equal length, each corresponding to \SI{400}{\ps} of data. Subsequently, the RDF is determined for each of these spatial and temporal blocks, resulting in RDFs with space- and time-resolved information on the degradation in our system. A demonstration of the partitioning into surface-specific regions and the remaining RDFs not treated in the main text (atom pairs: \ch{Pb}-\ch{Pb}, \ch{Pb}-\ch{Cs}, \ch{Pb}-\ch{I}, \ch{Cs}-\ch{Cs}, \ch{Cs}-\ch{I} and \ch{I}-\ch{I}) of the \ch{Pb}-rich and \ch{Pb}-poor (110) orthorhombic surface of \ch{CsPbI3} are shown in Figure~\ref{rdfs_combined}. 

\clearpage

\subsection{Defective surfaces}

\begin{figure*}[htbp!]
    \includegraphics{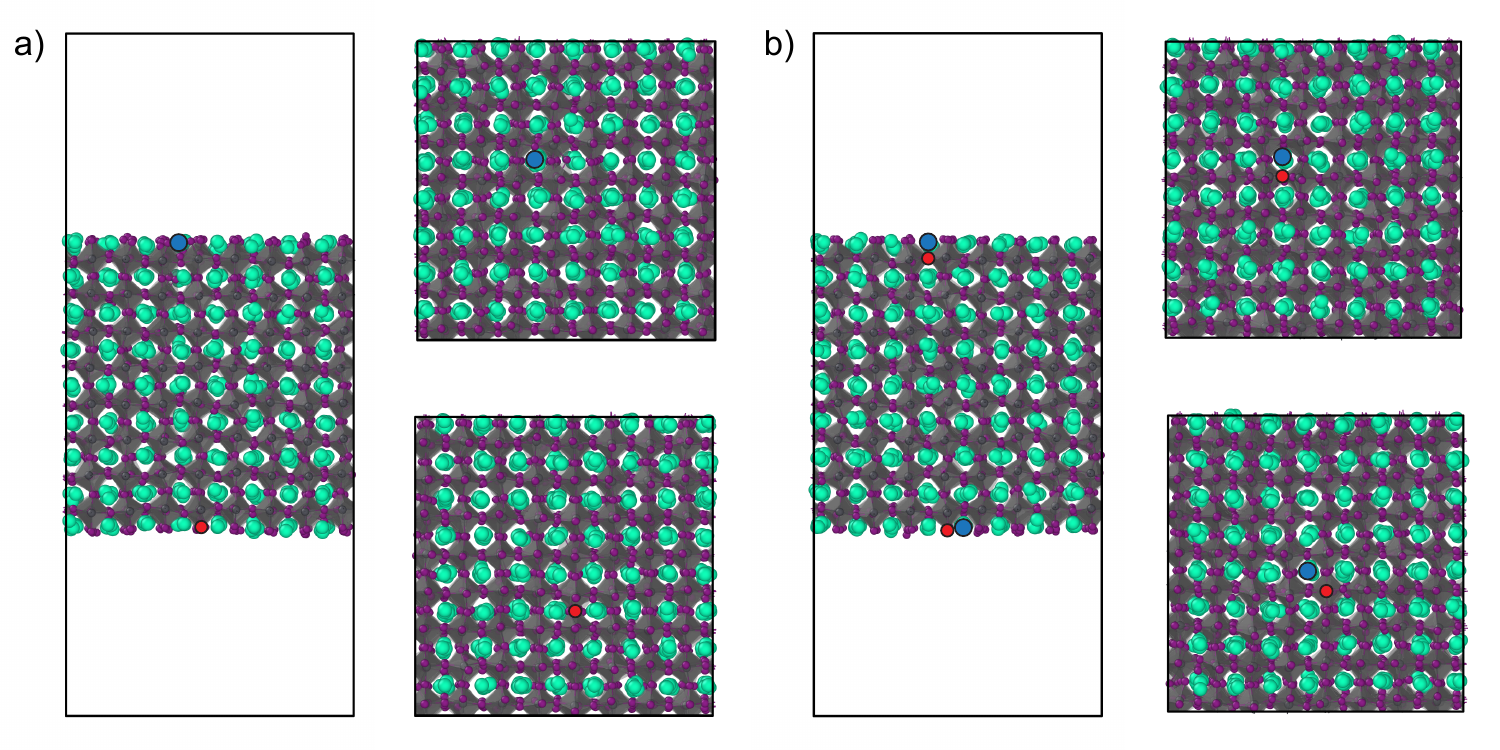}
    \caption{\label{defect_slabs} Structural models used to study the effect of point defects on the stability of the \ch{Pb}-poor (110) orthorhombic perovskite surface. The blue and red circles are used to indicate the position of the \ch{Cs} and \ch{I} vacancy, respectively. (a) A slab with a single \vacCs{} at the top surface and a single \vacI{} at the bottom surface. (b) A slab with closely-spaced defect pairs of \vacCs{} and \vacI{} at both the top and bottom surface.}
\end{figure*}

Defective perovskite surfaces are created with vacancies. The defects in these structural models are created by removing the atoms of certain species, thus resulting in vacancies in the lattice. For the isolated defects, we create slabs with a two vacancies in total, one of each kind. As shown in Figure~\ref{defect_slabs}a, the top surface has a \vacCs{} defect and the bottom surface a \vacI{} defect. For closely spaced pairs of \vacCs{} and \vacI{} defects, Figure~\ref{defect_slabs}b, we create the pair of defects at the top and bottom of the slab. The slabs in the simulations are at least \SI{50}{\angstrom} thick, with the lateral dimensions at least spanning \SI{50}{\angstrom} wide.

\clearpage

\subsection{Grain boundary evolution}

\begin{figure*}[htbp!]
    \includegraphics{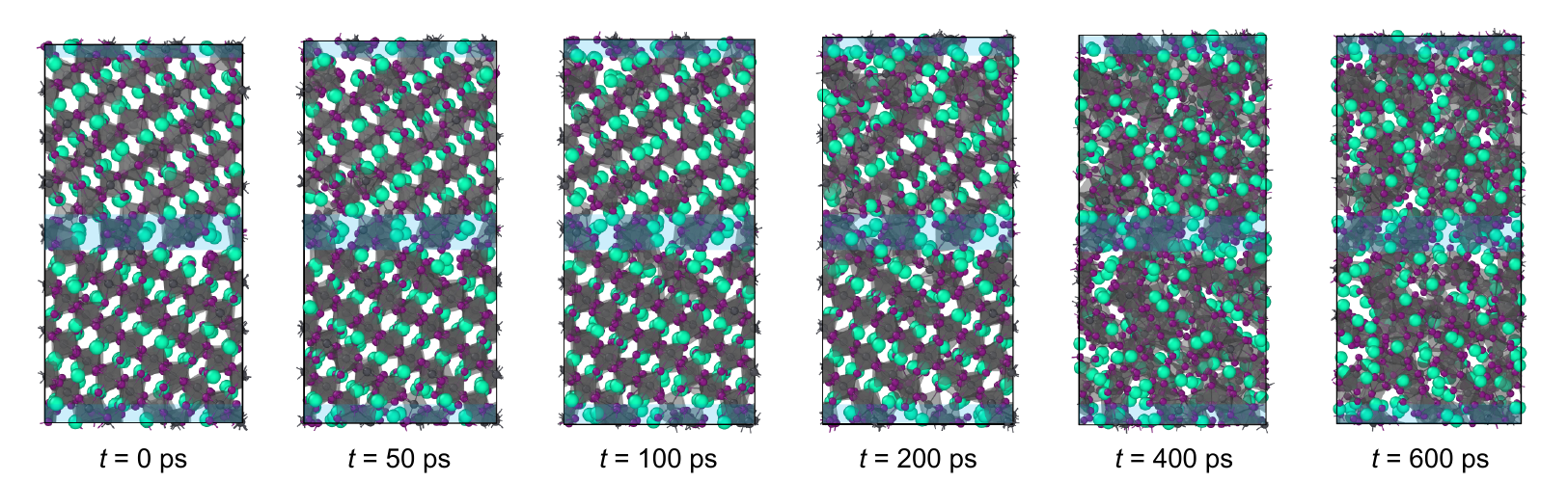}
    \caption{\label{3s112_evolution} Degradation of a \GBfirst{} grain boundary. The blue areas highlight the grain boundaries in the structures.}
\end{figure*}

\begin{figure*}[htbp!]
    \includegraphics{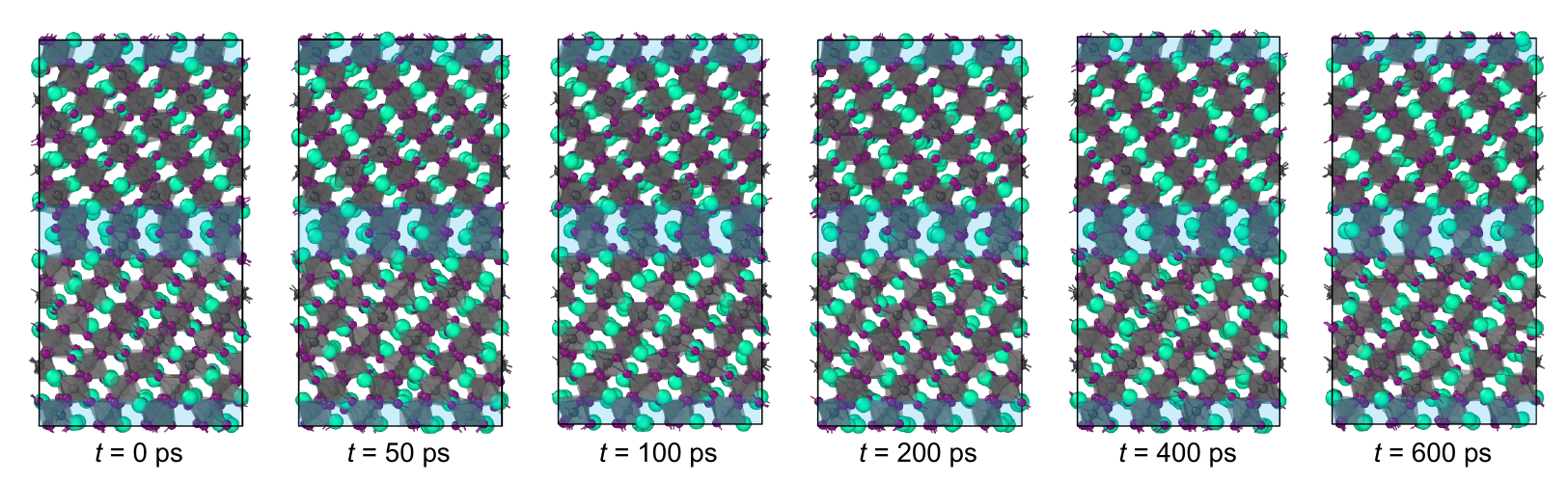}
    \caption{\label{3s111_evolution} Dynamical evolution of a \GBthird{} grain boundary. The blue areas highlight the grain boundaries in the structures.}
\end{figure*}

The degradation of the \GBsecond{} grain boundary using structural snapshots in the time interval ranging from \SI{0}{\ps} to \SI{600}{\ps} is explored in Figure 7. For a complete overview of all studied grain boundaries, the degradation of the \GBfirst{} grain boundary (Figure~\ref{3s112_evolution}) and dynamical evolution of the \GBthird{} grain boundary (Figure~\ref{3s111_evolution}) are shown here.

\clearpage

\bibliography{supporting}